\documentclass[preprint,pra,showpacs]{revtex4}
\usepackage{amssymb,amsmath}
\usepackage{graphicx}
\usepackage{epstopdf}
\usepackage{bm}
 \begin{document}
 
\title{Vortex dynamics in spin-orbit coupled Bose-Einstein condensates}
\author { Alexander L.\ Fetter}
\affiliation {Departments of Physics and Applied Physics, Stanford University,
   Stanford, CA 94305-4045, USA}

\date{\today}

\begin {abstract}
I use a time-dependent Lagrangian formalism and a variational trial function to study the dynamics of a two-component vortex in a spin-orbit coupled Bose-Einstein condensate (BEC).  For a single-component BEC, various experiments have validated this theoretical approach, for example a thermal quench  that yields a quantized vortex in roughly 25\% of trials~\cite{Frei10}.  To be definite, I assume the specific spin-orbit form used by Lin {\it et al.}~\cite{Lin09,Lin11} in recent NIST experiments, which introduces a spatial asymmetry because of the external Raman laser beams.    I here generalize this formalism to include a two-component order parameter that has quantized circulation in each component but not necessarily with the same circulation.  For example a singly quantized vortex in just one component yields a BEC analog of the half-quantized vortex familiar in $^3$He-A and in  $p$-wave chiral superconductors.  This and other unusual two-component vortices have both periodic trajectories and unbounded trajectories that leave the condensate, depending on the initial conditions. The optimized phase of the order parameter induces a term in the particle current that cancels the contribution from the vector potential, leaving  pure circulating current around the vortex.

\end{abstract}
\maketitle

\section{Introduction}

The remarkable creation  of quantized vortices in a gas of $^{87}$Rb atoms without external rotation of the condensate~\cite{Lin09} has stimulated much theoretical and experimental activity.  Previously, most  vortex experiments in ultracold dilute Bose-Einstein condensates (BECs)  relied on rotation of the condensate~\cite{Coop08,Fett09}, 
which is the analog of the rotating bucket generally used for both superfluid $^{4}$He and superfluid $^{3}$He.    

 In the past decade, however, a different picture of a rotating system has emerged. This new view arises from the  details of the well-known  transformation of the Hamiltonian to a rotating frame  $H \to H' = H -\bm\Omega \cdot \bm  L = H - \bm\Omega\times \bm r\cdot \bm p$, where $\bm \Omega$ is the angular velocity of the rotating frame.  It yields a Schr{\"o}dinger equation similar  to that for a charged particle in a uniform magnetic field.   Specifically, the rotation induces a synthetic (or artificial) vector potential with an effective ``symmetric'' gauge field $\bm A = M \bm \Omega \times \bm r$, where $M$ is the particle mass (see~\cite{Dali11} for a recent review of this exciting field).

 When a particle with charge $q$  travels from $\bm r_1$ to $\bm r_2$ in the presence of a vector potential $\bm A(\bm r)$, its wave function acquires a phase $S = (q/\hbar)\int_1^2 \bm A(\bm r')\cdot d\bm r'$.  If by some means, one can create such a phase, even a neutral atom can experience a synthetic gauge field.  Hence much of the recent work on cold atoms has focused on phase engineering of the quantum wave function, particularly the possibility of studying many-body systems~\cite{Bloc08} that  may not be readily accessible in conventional condensed-matter materials.  
 
 Spielman and his group have used Raman-induced transitions and a magnetic field gradient to generate such a spatially varying synthetic vector potential $A_x\propto y$ that acts like ``Landau'' gauge and yields a nearly uniform synthetic magnetic field (equivalent to the uniform rotation of most previous  experiments).  In this way, they  created a system of quantized vortices at rest in the laboratory frame~\cite{Lin09}.  They subsequently made  a form of spin-orbit coupling in a two-component BEC~\cite{Lin11} with the synthetic gauge field $\bm A =A_x\, \bm{\hat x}\,\sigma^z$ where $\sigma^z$ is a   Pauli matrix that acts on the two components of the state vector (see~\cite{Gali13} for a recent review).
 
 Radi{\'c} {\it et al.}~\cite{Radi11} studied the question of vortex formation in a spin-orbit coupled BEC.  They  point out the difficulty of stirring such a condensate by the familiar technique of external  rotation, which would involve rotating not only the trap but also the Raman laser beams (and perhaps also the external magnetic field).  Stimulated by these concerns, I here explore the dynamics of a 
 vortex in such a two-component spin-orbit coupled BEC that is  stationary in the laboratory frame and hence nonrotating.

In an elegant experiment, Freilich {\it et al.}~\cite{Frei10} performed rapid thermal quenches of a one-component gas of $^{87}$Rb atoms  deep into the superfluid  BEC regime.  Roughly 25\% of the time, they  created  a vortex  with random $\pm$ orientation of the circulation.   To monitor the vortex dynamics in real time, they applied a short microwave pulse to transfer about 5\% of the condensate atoms to an untrapped state that falls and expands, making visible the position of the vortex as a hole in the condensate density. Repeating this process up to $\sim$ 8 times yields  time-lapse pictures of the vortex motion. The resulting sequence of images shows clearly the  position of the moving vortex.    This method of vortex creation is important because it does not rely on rotation of the condensate (see~\cite{Neel10} for a  different valuable nonrotating approach to the dynamics of a vortex dipole/pair).  In principle this technique~\cite{Frei10}   could  also serve to study vortex dynamics in a spin-orbit coupled BEC.  Note, however, that  the presence of a Zeeman magnetic field~\cite{Lin11} would necessitate a laser dipole trap and a modified approach to releasing a sequence of  small samples to image the vortex motion~\cite{Rama12}.

If it is indeed possible to create a vortex in a nonrotating spin-orbit coupled BEC, each component must have  single-valued circulation.
 There is no obvious requirement that the two components have the same circulation, even though such states may have higher energy.  This study seeks  observable features of the vortex motion that would identify such unusual states with different circulation in the two components. One possibility  would be unit circulation in one component and zero circulation in the other, which would be an analog of the ``half-quantum'' vortex predicted to occur in superfluid $^3$He-A films~\cite{Volo77,Salo85} and in similar chiral  $p$-wave ($p_x \pm ip_y$) superconductors such as Sr$_2$RuO$_4$~\cite{Kall12}.

Section II reviews the time-dependent variational Lagrangian treatment of vortex dynamics in a one-component BEC. 
  Section III then generalizes this treatment of vortex dynamics to the two-component single-particle Hamiltonian introduced in~\cite{Spie09} and implemented experimentally in~\cite{Lin11}.

\section{Review of vortex dynamics in a single-component BEC}
 
 The motion of a vortex in a single-component trapped BEC arises from the presence of the trap potential $V_{\rm tr}(\bm r)$.  In the usual case of an axially symmetric harmonic  trap with $V_{\rm tr} = \frac{1}{2}M(\omega_\perp^2 r^2 + \omega_z^2 z^2)$, the radial force $-\partial V_{\rm tr}(r,z)/\partial r$ is linear and points inward.  The intrinsic angular momentum of the vortex  causes it to act like a gyroscope, and it moves  perpendicular to the radial force with a velocity $\propto \hat{\bm z}\times \bm \nabla_\perp V_{\rm tr}$~\cite{Fett09} (this effect is sometimes called the ``Magnus'' force).  As a result, the vortex  precesses with uniform circular motion in the same direction as its circulating velocity.  

The time-dependent variational Lagrangian method provides a convenient way to analyze this precessional motion (see~\cite{Fett09}, Sec.\ III.B.2.a for a brief introduction). 
As usual, I introduce a condensate wave function $\Psi(\bm r,t)$ to characterize the low-temperature  Bose-Einstein condensate.  It obeys the  time-dependent Gross-Pitaevskii (GP) equation
\begin{equation}\label{GP}
i\hbar\frac{\partial \Psi}{\partial t} = \left(-\frac{\hbar^2\nabla^2}{2M} + V_{\rm tr} +V_{\rm H}\right) \Psi.
\end{equation}
In essence, this  nonlinear Schr{\"o}dinger equation includes an effective Hartree potential $V_{\rm H}(\bm r) = gn(\bm r) = g|\Psi(\bm r)|^2$ arising from the local interaction with the other particles.  Here, $g = 4\pi \hbar^2 a/M$ is an interaction constant, and $a\ (\sim$ a few nm) is the $s$-wave scattering length.

The essential observation is that  Eq.\ (\ref{GP}) is the Euler-Lagrange equation for the time-dependent Lagrangian functional
\begin{equation}\label{L}
L[\Psi] = T[\Psi]- E_{\rm GP}[\Psi],
\end{equation}
where 
\begin{equation}\label{T}
T[\Psi] =  \int dV \, \frac{i\hbar}{2}\left(\Psi^*\frac{\partial \Psi}{\partial t} -\frac{\partial \Psi^*}{\partial t}\Psi\right) 
\end{equation}
plays the role of the conventional kinetic energy and 
\begin{equation}\label{EGP}
E_{\rm GP}[\Psi] = \int dV \left( \frac{\hbar^2}{2M}\, |\bm \nabla\Psi|^2 + V_{\rm tr}\,|\Psi|^2 + \frac{g}{2} \,|\Psi|^4 \right)
\end{equation}
is the GP energy functional for a nonrotating condensate (it plays the role of the potential energy).  Minimization of $E_{\rm GP}$ with fixed normalization ($\int dV\,|\Psi|^2 = N $) yields the usual time-independent GP equation.  If the condensate wave function depends on one or more parameters, the resulting Lagrangian functional yields approximate Lagrangian equations of motion for these parameters.  

Initially, this Lagrangian approach served  to  study the monopole and quadrupole collective-mode  frequencies of a  trapped condensate~\cite{Pere96}, but it soon yielded valuable estimates of the vortex precession frequency in a harmonic trap~\cite{Lund00,Svid00}.  Here, the position $\bm r_0$ of the vortex is the variational parameter, and the resulting Lagrangian $L(\dot{\bm r}_0,\bm r_0)$ yields dynamical equations for the vortex motion.
 
For simplicity, I consider a two-dimensional condensate with $N$ particles and tight Gaussian confinement in the $z$ direction with  small oscillator length $d_z=\sqrt{\hbar/(M\omega_z)}$.  In the Thomas-Fermi (TF)  limit, I assume a two-dimensional  condensate wave function 
\begin{equation}\label{Psi}
\Psi(\bm r) = \sqrt{n_0}\,e^{iS}\,\left( 1-\frac{r^2}{R^2}\right)^{1/2},
\end{equation}
where $R$ is the TF condensate radius, $S$ is the phase, and the two-dimensional central density $n_0 = 2N/(\pi R^2)$ follows from the normalization condition.  In this two-dimensional situation, the GP energy functional in Eq.\ (\ref{EGP}) involves a two-dimensional integral  $d^2 r$ and has an effective  two-dimensional interaction constant $g_{2d} = g/(\sqrt{2\pi}\,d_z)$. 

Assume a singly quantized vortex at $\bm r_0$, leading to a phase singularity
\begin{equation}\label{phase}
S = \arctan\left(\frac{y-y_0}{x-x_0}\right).
\end{equation}
This choice of  phase  means that  the trial wave function depends on the position of the vortex, which serves as a parameter in the variational Lagrangian formalism.   The trap energy and the interaction energy depend only on the radial profile, and  straightforward integrations yield
\begin{equation}\label{int-trap}
 E_{\rm trap} + E_{\rm int}= \frac{1}{6} M\omega_\perp^2 R^2 N + \frac{8N^2}{3\sqrt{2\pi}}\frac{\hbar^2 a}{MR^2 d_z}.
\end{equation}
Minimizing this quantity with respect to $R^2$ yields the equilibrium condensate radius
\begin{equation}\label{R}
R^4 = \frac{16}{\sqrt{2\pi}}\frac{Nad_\perp^4}{d_z},
\end{equation}
where $d_\perp = \sqrt{\hbar/(M\omega_\perp)}$ is the transverse oscillator length.

In the Thomas-Fermi limit, the slow spatial variation of the density means that the kinetic energy arises only from the velocity (the gradient of the phase)
\begin{equation}
\bm v = \frac{\hbar}{M}\bm \nabla S =\frac{\hbar}{M} \frac{\bm{\hat z}\times(\bm r-\bm r_0)}{|\bm r-\bm r_0|^2};
\end{equation}  
it exhibits circulating flow around the vortex.  I have not included an image vortex, since it makes only a small contribution the energy [note that the current $\bm j = n\bm v$ vanishes at the boundary since $n(r$) vanishes there].  The integration for the kinetic energy is most simply evaluated with a stream function $\chi(\bm r,\bm r_0)  = \ln|\bm r-\bm r_0|$ (see Appendix  A of~\cite{Kim04}).  This calculation is analogous to the evaluation of the energy of a quantized superconducting flux line in the London limit~\cite{deGe66}, but the details are a bit more intricate because of the  nonuniform TF density profile.  A detailed analysis gives
\begin{equation}\label{KE0}
 E_k = \frac{\hbar^2N}{2MR^2} \left\{\left(1-u_0^2\right) \left[2\ln\left(\frac{R}{\xi}\right) + \ln\left(1-u_0^2\right) 
\right] + 2u_0^2 -1\right\},
\end{equation}
where $u_0 = r_0/R$ is the dimensionless radial position of the vortex  and $\xi =\hbar/\sqrt{2n_0g_{\rm 2d} } $ is the healing length (typically $\xi \sim$ a fraction of 1 $ \mu$m).

The time-dependent part $T[\Psi]$ of the Lagrangian is readily evaluated to give
\begin{equation}\label{T1}
T =-\hbar N  \left(u_0^2-\frac{1}{2}u_0^4\right) \dot\phi_0,
\end{equation}
which involves the angular velocity $\dot\phi_0$ of the vortex (but not the radial velocity $\dot u_0$).  In this  single-component situation,  the Lagrangian  for the vortex dynamics depends only on the angular velocity $\dot\phi_0$ and the radial position $u_0$:  $L (\dot\phi_0, u_0)= T(\dot\phi_0,u_0)- E_k( u_0) $.   The corresponding Euler-Lagrange equations simplify considerably to give equations with a formal  Hamiltonian structure
\begin{eqnarray}\label{dyn}
\dot\phi_0 & = & -\frac{1}{2N\hbar u_0(1-u_0^2)}\frac{\partial E_{k}}{\partial u_0},\\[.3cm]
\dot u_0& =& \frac{1}{2N\hbar u_0(1-u_0^2)}\frac{\partial E_{k}}{\partial \phi_0}= 0\nonumber ,
\end{eqnarray}
and the second equation here vanishes because $E_k$ does not  depend on $\phi_0$.

In the present case of a single-component two-dimensional  condensate with no dissipation, the energy is conserved, and  the radial position $u_0$ remains fixed since the energy depends only on $u_0$. 
 Equation (\ref{dyn}) shows that   the vortex precesses uniformly at a rate 
\begin{equation}
\dot{\phi}_0 = \frac{\hbar}{MR^2} \frac{1}{1-u_0^2}\left[\ln\left(\frac{R}{\xi}\right)+\frac{1}{2}\ln\left(1-u_0^2\right)-\frac{1}{2}\right].
\end{equation}
The generalization of this analysis  to a three-dimensional disk-shaped condensate~\cite{Lund00,Svid00} agrees well with recent experimental observations following a thermal quench~\cite{Frei10} and also with earlier ones that used Rabi coupling and a stirring laser~\cite{Ande00}.

\section{spin-orbit coupled two-component condensate}

The generalization to a two-component spin-orbit coupled BEC is relatively straightforward, but it is important to emphasize that I here use the particular  spin-orbit Hamiltonian  recently implemented experimentally by Spielman and his group~\cite{Spie09,Lin09,Lin11}.  There are alternative more symmetric versions (see, for example,~\cite{Zhai11,Camp11,Afta13} for a discussion of these ``Rashba'' coupling schemes), but the present choice has the advantage of being realistic for future experiments involving vortex dynamics.

\subsection{Evaluation  of energy}

 For a spin-orbit coupled BEC, the most significant alteration is in the GP energy functional, where the single-particle Hamiltonian  has the more general $2\times 2$ matrix  structure 
\begin{equation}\label{H0}
{\cal H}_0 = \frac{\hbar^2}{2M}\left(-i\bm \nabla \sigma^0+ k_0\,\bm {\hat \bm x}\,\sigma^z\right)^2 + \frac{\hbar\delta}{2 }\sigma^z+ \frac{\hbar\Omega}{2}\sigma^x + V_{\rm tr}\,\sigma^0.
\end{equation}
Here, $\sigma^j$ for $j=1,2,3$ is one of the Pauli matrices and $\sigma^0$ is the unit matrix.  
As discussed in~\cite{Spie09,Lin09,Lin11}, this spinor Hamiltonian has three new parameters under experimental  control:  $k_0$ is the wavenumber of the Raman laser beams, $\Omega$ is the associated  Rabi frequency, related to the intensity of the Raman laser beams, and $\delta$ is the detuning, controlled by the external magnetic field. Note the presence of  the uniform synthetic gauge field $\bm A = -\hbar k_0\, \hat{\bm x}\,\sigma^z$ proportional to $\sigma^z$. To be very specific, this particular form of the spin-orbit Hamiltonian appears implicitly in the Methods section of~\cite{Lin11} below Eq.~(2), before the cyclic  global pseudo-spin rotation $\sigma^z\to\sigma^y, \sigma^y\to\sigma^x, \sigma^x\to \sigma^z$ used to obtain their  final Hamiltonian $H_2$ (see also~\cite{Lu13} for a related discussion). 

The Lagrangian retains the form seen in Eq.~(\ref{L}), but the energy functional  now  includes the single-particle spin-orbit Hamiltonian from Eq.~(\ref{H0})
\begin{equation}\label{EGPSO}
E_{\rm GPSO}[\Psi] = \int d^2r\left( \Psi^\dagger {\cal H}_0\Psi + {\textstyle
\frac{1}{2}}g_{2d}|\Psi^\dagger\Psi|^2\right).
\end{equation}
Here I do not include the subtle  effect of the interactions discussed in~\cite{Lin11,Lu13}.  They lead to phase separation for increasing Rabi frequency $\Omega$ above $\Omega/E_R\gtrsim 0.2$, where $E_R = \hbar^2k_0^2/(2M)$  is the recoil energy acquired by an atom on absorbing a single Raman photon with wave number $k_0$.

To account for the spinor character of the trial wave function, I generalize Eq.~(\ref{Psi})  to write 
\begin{equation}\label{Psi2}
\Psi(\bm r) = \sqrt{n_0} \left(1-\frac{r^2}{R^2}\right)^{1/2} \zeta,
\end{equation}
where 
\begin{equation}\label{zeta}
\zeta = \left(
\begin{matrix}  e^{iS_1} \cos(\frac{1}{2}\chi)\\
e^{iS_2}e^{i\eta} \sin(\frac{1}{2}\chi)
\end{matrix}\right) 
\end{equation}
is a normalized two-component spinor with constant parameters $\chi$ and $\eta$.    I assume a vortex located at $\bm r_0$,  and the phases 
\begin{equation}\label{phase2}
S_j = m_j \arctan\left(\frac{y-y_0}{x-x_0}\right) + \alpha x
\end{equation}
(with $j = 1,2$) contain the vortex contribution seen in Eq.~(\ref{phase}) along with an additional velocity contribution $\alpha x$ to account for the spatial asymmetry of ${\cal H}_0$  seen in Eq.~(\ref{H0}).  Here, the vortex winding numbers $m_1$ and $m_2$ must be integers to make the  spinor wave function single-valued, but there seems no essential reason for them to be equal.  Specifically, I consider the cases $m_1 =1$ and $m_2 = 1, 0, -1$ as illustrative of various possibilities.
 
The single-particle Hamiltonian ${\cal H}_0$ in Eq.~(\ref{H0}) contains several Pauli matrices, and their expectation values are readily calculated:  $\zeta^\dagger \sigma^0\zeta = 1$ (normalization), $\zeta^\dagger\sigma^z\zeta = \cos \chi$ and 
 \begin{equation}\label{sigmax}
\zeta^\dagger \sigma^x\zeta = \cos(S_1-S_2-\eta)\sin\chi.
\end{equation}
With the trial function (\ref{Psi2}),  the trap energy and  interaction energy remain unchanged and yield the values given in Eq.~(\ref{int-trap}) with the  TF radius $R$ from Eq.~(\ref{R}).
 
 In the Thomas-Fermi approximation, the kinetic energy arises only from the gradient of the phase and yields 
 \begin{equation}\label{KE}
E_k= \frac{\hbar^2n_0}{2M}\,\int d^2 r\left(1-\frac{r^2}{R^2}\right) \frac{1}{2} \left[ \left(\bm\nabla S_1+ k_0\hat{\bm x}\right)^2 (1+\cos\chi) + \left(\bm\nabla S_2- k_0\hat{\bm x}\right)^2 (1-\cos\chi) \right].
\end{equation} 
A detailed analysis yields $E_{k0} + E_{kv}$, where 
\begin{equation}\label{Ek0}
E_{k0} = \frac{\hbar^2 N}{2M}\left(\alpha ^2 + 2\alpha k_0 \cos\chi +k_0^2\right)
\end{equation}
is the kinetic  energy in the absence of a vortex and 
\begin{eqnarray}\label{Ekv}
E_{kv} &=&   \frac{\hbar^2 N}{MR^2}r_0\left(1-\frac{r_0^2}{2R^2}\right) \left\{\alpha \sin\phi_0[m_1+m_2 +(m_1-m_2)\cos\chi]\right.\nonumber\\[.2cm]
& +& \left.k_0\sin\phi_0[m_1-m_2 +(m_1+m_2)\cos\chi]\right\}\nonumber\\[.2cm]
& +& \frac{N\hbar^2}{2MR^2} \left\{ \left(1 -\frac{r_0^2}{R^2}\right)\left[2\log\left(\frac{R}{\xi}\right) + \log\left(1-\frac{r_0^2}{R^2}\right)\right] + 2\frac{r_0^2}{R^2} -1\right\}
\nonumber \\[.2cm]
&\times& \left[m_1^2 + m_2^2 +(m_1^2 -m_2^2)\cos\chi\right]
\end{eqnarray}
is the additional kinetic energy associated with the presence of a vortex at $\bm r_0$.

In the simplest situation without a vortex, the remaining spin-orbit energy becomes
\begin{equation}\label{ESO}
\left(E_{SO}\right)_0= {\textstyle\frac{1}{2}}\hbar N\left(\delta\cos\chi + \Omega\cos\eta\sin\chi\right).
\end{equation}
My strategy is to minimize the energy with respect to the parameters $\alpha$, $\eta$ and $\chi$ assuming there is no vortex, and then retain the resulting values for the case of a vortex. As justification, note that  the recoil energy $E_R=\hbar^2k_0^2/(2M)$ sets the basic laser energy scale.  In contrast,  the vortex energy [the last term in Eq.~(\ref{Ekv})] is of order $\hbar^2\ln(R/\xi)/(MR^2)$, which is smaller by a factor $\ln(R/\xi)/(k_0R)^2\ll 1$ (note that $k_0R\sim 10-40$ for typical parameters).  Thus the vortex energy has only a small effect on this minimization.  More generally,  this  variational analysis yields an optimal description of vortex dynamics within the imposed trial wave functions. Keeping  only Eq.~(\ref{Ek0}),  the minimization with respect to $\alpha$  yields the equilibrium value 
\begin{equation}\label{alpha}
\alpha = -k_0\cos\chi. 
\end{equation}
 Similarly, minimization with respect to $\eta $ gives the condition $\sin\eta =0$, and the appropriate choice is $\eta=\pi$, so that $\cos\eta = -1$. 

For these values, the relevant vortex-free energy is $E_{k0} +(E_{SO})_0 = NE_R\sin^2\chi -\frac{1}{2} N\hbar\Omega \sin\chi$, where  I set $\delta = 0$ for simplicity.  The minimization with respect to $\chi$ yields 
\begin{eqnarray}\label{chi}
\sin \chi = \begin{cases}
			\hbar\Omega/(4E_R) &\text{if $0\le \hbar\Omega\le 4E_R$;}\\
			1 &\text{ otherwise.}
			\end{cases}
\end{eqnarray}
To ensure miscibility of the two components~\cite{Lin11,Lu13}, I typically consider only the range $\hbar\Omega/E_R \lesssim 0.2$, so that $\sin \chi$ remains small.

It is convenient to use $\hbar^2 N/(2MR^2)$ as the unit of energy with the dimensionless variable $u_0 = r_0/R$ and to set $\alpha=-k_0\cos\chi$ explicitly.  Hence the dimensionless vortex energy in Eqs.~(\ref{Ek0}) and (\ref{Ekv}) becomes 
\begin{eqnarray}\label{Ek}
\tilde E_{k} &=&  k_0^2 R^2\sin^2\chi  +2u_0(1-{\textstyle\frac{1}{2}}u_0^2)\,
 k_0R \sin^2\chi\,(m_1-m_2) \sin\phi_0\nonumber\\[.2cm]
& +& \left\{ (1 -u_0^2)[2\log(R/\xi ) + \log(1-u_0^2)] + 2u_0^2 -1\right\}
\nonumber \\[.2cm]
&\times& \left[m_1^2 + m_2^2 +(m_1^2 -m_2^2)\cos\chi\right].
\end{eqnarray}

The remaining contribution to the energy functional  arises from the Rabi coupling.  With $\delta = 0$, this term involves the integral of $-\cos(S_1-S_2) = -{\rm Re}\,e^{i(S_1-S_2)}$ averaged over the Thomas-Fermi density profile $n_{\rm TF}({u})= n_0(1-u^2)$, where $u = r/R$.  This complex phase factor  can be rewritten as 
\begin{equation}\label{arg}
e^{i(S_1-S_2)} = \left(\frac{z-z_0}{|z-z_0|}\right)^m,
\end{equation}
where $m = m_1-m_2$, $z= x+iy= re^{i\phi}$ is a complex variable, and $z_0 = x_0 + i y_0 = r_0e^{i\phi_0}$ denotes the two-dimensional position of the  vortex.  I introduce  the spatial integral 
\begin{equation}\label{fm}
f_m(u_0)\,e^{im\phi_0}=\frac{2}{\pi R^2}\int d^2 r\left(1-\frac{r^2}{R^2}\right) \left(\frac{z-z_0}{|z-z_0|}\right)^m,
\end{equation}
where $u_0 = r_0/R$ and $f_m(u_0)$ is real.  By inspection, $f_0(u_0) = 1$.

The case $m=1$ is nontrivial, because the angular integral leads to complete elliptic integrals $K$ and $E$~\cite{Grad65,Dwig68}
\begin{eqnarray}\label{ang}
f_1(u_0) &=& \frac{2}{\pi} \int_0^1u\,du\,(1-u^2)\int_{-\pi}^\pi d\phi\,\frac{u\,\cos\phi-u_0}{\sqrt{1-2u\cos\phi + u^2}}\nonumber\\[.2cm]
&=& \frac{8}{\pi}  \int_0^1u\,du\,(1-u^2)\left\{ -\theta(u_0-u) E\left(\frac{u}{u_0}\right) \right.\nonumber \\[.2cm]
&+& \left. \theta(u-u_0)\left[\left(\frac{u}{u_0}-\frac{u_0}{u}\right)K\left(\frac{u_0}{u}\right) -\frac{u}{u_0}E\left(\frac{u_0}{u}\right)\right]\right\},
\end{eqnarray}
where $\theta$ denotes the unit positive step function. The remaining radial integral has an exact expression in terms of  various generalized hypergeometric functions, but the result is unwieldy.  For most purposes, I  therefore use the following  accurate  analytic 
expression 
\begin{equation}\label{f1}
f_1(u_0)\approx -{\textstyle\frac{4}{3}} u_0 + cu_0^3,
\end{equation}
where $c =\frac{4}{3} -128/(45\pi)\approx 0.4279$.  This approximation reproduces both the slope $f_1'(0)= - \frac{4}{3}$ at $u_0=0$ and the value $f_1(1)= -128/(45\pi)\approx- 0.9054$ at $u_0=1$.

The similar evaluation of $f_2(u_0)$ is straightforward and yields
\begin{equation}\label{f2}
f_2(u_0) = u_0^2-\textstyle\frac{1}{3}u_0^4.
\end{equation}
As a result, the corresponding dimensionless spin-orbit energy becomes 
\begin{equation}\label{ESO2}
\tilde E_{\rm SO}= -\frac{MR^2\Omega}{\hbar}\sin\chi\cos(m\phi_0) \,f_m(u_0),
\end{equation}
where $m = |m_1-m_2|$ is an integer (I here consider the cases $m = 0, 1,2$).

In the context of vortex dynamics,  the only relevant  parts of the GP energy $E_{\rm GP} $ are those that depend on the coordinates of the vortex.  Specifically, I here include Eqs.~(\ref{Ek}) and (\ref{ESO2}), whose sum
\begin{equation}\label{Ev}
\tilde E_{ v} = \tilde E_{k} + \tilde E_{\rm SO}
\end{equation} determines the effective energy in the dynamical Lagrangian.  In contrast to the case of a single-component vortex
where Eq.~(\ref{KE0}) depends only on the radial coordinate, the effective energy $\tilde E_v$ here depends explicitly on the angular position of the vortex through the factors $\sin\phi_0$ and $\cos(m\phi_0)$.  Note that this dependence arises from the anisotropy of the particular single-particle Hamiltonian~\cite{Lin11} used by Spielman and his group.  Presumably, such anisotropy would not appear for  pure Rashba coupling~\cite{Zhai11,Camp11,Afta13}, which is isotropic in the $xy$ plane.

\subsection{Vortex dynamics}

The time-dependent part  of the Lagrangian $\tilde T$ determines the vortex dynamics through the time dependence of the phases $S_j$.  A direct analysis yields a dimensionless  two-component generalization of Eq.~(\ref{T1}):
\begin{equation}\label{T2}
\tilde T = -\frac{2MR^2\dot{\phi}_0}{\hbar}\left(u_0^2-\frac{1}{2}u_0^4\right)\left[m_1+m_2 +\left(m_1-m_2\right)\cos\chi\right],
\end{equation}
where I 
 use the same dimensionless variables as in Eq.~(\ref{Ek}).

The  Lagrangian $\tilde L = \tilde T - \tilde E_v$ depends on the dimensionless variables $u_0, \phi_0$ and $\dot{\phi}_0$ but not on the radial velocity $\dot{u}_0$.  Hence the Euler-Lagrange equation for the variable $u_0$ simplifies and yields $\partial \tilde L /\partial u_0 = 0 $.  An easy calculation gives
\begin{equation}\label{phidot}
\dot{\phi}_0 =- \frac{\hbar}{4MR^2}\frac{1}{u_0(1-u_0^2)[m_1+m_2+(m_1-m_2)\cos\chi]}\,\frac{\partial \tilde E_v}{\partial u_0},
\end{equation}
which generalizes Eq.~(\ref{dyn}) to the two-component spin-orbit coupled vortex.  Note that $\tilde E_v$ depends on both $u_0$ and $\phi_0$, so that in general  the angular velocity of the vortex now  depends explicitly on the angular position of the vortex.

The corresponding Euler-Lagrange equation for the angular variable $\phi_0$ is a bit more complicated because $\tilde L$ depends  explicitly on $\dot{\phi}_0$ through $\tilde T$
\begin{equation}\label{phi0}
\frac{d}{dt}\left(\frac{\partial \tilde L}{\partial \dot{\phi}_0}\right) =\frac{\partial \tilde L}{\partial \phi_0}.
\end{equation}
A straightforward analysis yields 
\begin{equation}\label{udot}
\dot{u}_0 = \frac{\hbar}{4MR^2}\frac{1}{u_0(1-u_0^2)[m_1+m_2+(m_1-m_2)\cos\chi]}\,\frac{\partial \tilde E_v}{\partial \phi_0}.
\end{equation}
Equations (\ref{phidot}) and (\ref{udot}) exhibit a Hamiltonian structure, which has the following  important consequence.  As the vortex moves under these dynamical equations, the energy is conserved:
\begin{equation}\label{cons}
\frac{d\tilde E}{dt} = \frac{\partial \tilde E_v}{\partial u_0}\,\dot{u}_0 + \frac{\partial \tilde E_v}{\partial \phi_0}\,\dot{\phi}_0 = 0.
\end{equation}
Hence the motion of this two-component vortex follows a contour of constant energy $\tilde E_v$.

\begin{figure}[h]
 \begin{center}
  \includegraphics[width=2.5in]{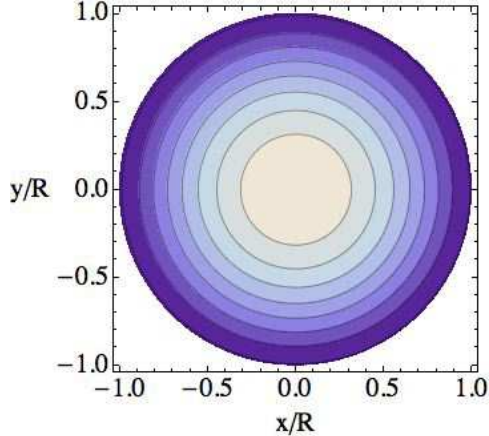}
  \end{center}
  \caption{(color online) Contours of constant dimensionless energy $\tilde E_v$ for the case of equal circulations $m_1=m_2=1$. The axes show dimensionless Cartesian coordinates, normalized by the Thomas-Fermi radius $R$. Here the circular contours  mean that the vortex precesses uniformly in accordance with Eq.~(\ref{phidot}).}
\end{figure}

Given the conservation of energy for this dynamical system, it is natural to focus on the contours of constant energy $\tilde E_v$.  For  a two-component spin-orbit coupled vortex, the simplest example has  equal winding numbers $m_1=m_2 = 1$, when only the kinetic energy in Eq.~(\ref{Ek}) is relevant.  The corresponding  circular energy  contours are essentially identical with those for a single component vortex [see Eq.~(\ref{KE0})].  Figure 1 shows the energy  contours for $m_1=m_2$, where the precession frequency $\dot{\phi}_0$ is proportional to the negative radial gradient (namely the density of contour lines).

\begin{figure}[ht]
 \begin{center}
  \includegraphics[width=3.5in]{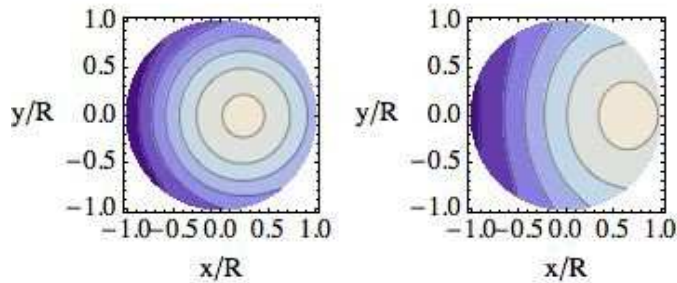}
\end{center}
  \caption{(color online) Contours of constant dimensionless energy $\tilde E_v$ in Eq.~(\ref{Ev})  for the case of different  circulations with $m=|m_1-m_2|=1$ (for example, $m_1=1, m_2=0$), which is analogous to a half-quantum vortex in superfluid $^3$He-A or in a $p$-wave chiral superconductor.  The axes show dimensionless Cartesian coordinates, normalized by the Thomas-Fermi radius $R$.  Energy surfaces  shown are  for (left) $\hbar\Omega/E_R = 0.1$ and (right) $\hbar\Omega/E_R = 0.2$.  Note the presence of both closed vortex trajectories  that remain in the condensate and open vortex trajectories that leave the condensate. Note also the dipole (lateral) displacement of the local maximum of $\tilde E_v$}
\end{figure}

\begin{figure}[h]
 \begin{center}
  \includegraphics[width=3.5in]{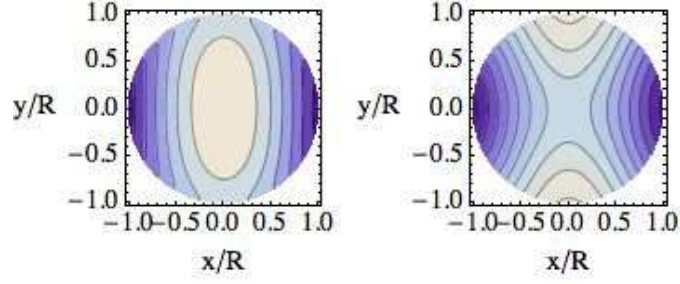}
\end{center}
  \caption{(color online) Contours of constant dimensionless energy $\tilde E_v$ in Eq.~(\ref{Ev})  for the case of different  circulations $m=|m_1-m_2|=2$ with, for example, $m_1=1, m_2=-1$.  The axes show dimensionless Cartesian coordinates, normalized by the Thomas-Fermi radius $R$.  Energy surfaces   shown are for (left) $\hbar\Omega/E_R = 0.1$, where trajectories  near the center of the condensate are closed and remain in the condensate, and (right)  $\hbar\Omega/E_R = 0.2$, where most of the trajectories  are  open and leave the condensate.   Note the manifest quadrupole distortion  of $\tilde E_v$.}
\end{figure}

The behavior is more unusual for a half-quantum vortex with (for example) $m_1=1$ and $m_2=0$. The corresponding energy (\ref{Ev}) now contains contributions that depend on the angular position $\phi_0$ of the vortex.  To understand the vortex trajectories, consider the relevant  surfaces of constant energy (Fig.\ 2), which now involve a displaced maximum because of the dipole terms in $\tilde E_v$  proportional to $\sin\phi_0$ and $\cos\phi_0$.  Evidently, even a small value of the Rabi frequency $\Omega$ yields some unbound trajectories, and increasing $\Omega$ leads to more unbound trajectories.

A similar situation occurs for the case of $m=2$, for example $m_1=1$ and $m_2=-1$, where $\tilde E_{\rm SO} $ contains the quadrupole distortion $\cos(2\phi_0)$.  Figure 3 shows the corresponding contours of constant energy.  For relatively small $\hbar\Omega/E_R=0.1$, both closed and open trajectories exist;  in contrast,  for larger $\hbar  \Omega/E_R = 0.2$, most trajectories are open and leave the condensate.  

\subsection{Particle current}

The two-component single particle Hamiltonian in Eq.~(\ref{H0}) for a spin-orbit coupled BEC readily yields a time-dependent Gross-Pitaevskii equation including the Hartree nonlinear interaction (here assumed to be independent of the spin index for simplicity). It is straightforward to obtain a conservation law $\partial n/\partial t + \bm \nabla\cdot \bm j = 0$ for the particle density  $n=\Psi^\dagger\Psi$, with the particle-current density (see, for example~\cite{Ozaw12})
\begin{equation}\label{j}
\bm j = \frac{\hbar}{2Mi}\left[\Psi^\dagger\bm \nabla \Psi -(\bm \nabla \Psi^\dagger)\Psi\right] -\frac{1}{M}\Psi^\dagger\bm A\Psi.
\end{equation}
The first term is the usual phase contribution, and the second arises from the gauge field $\bm A = -\hbar k_0\,\hat{\bm x} \,\sigma^z$.  This form of $\bm j$ follows directly as the expectation value of the velocity $\bm v = (\bm p - \bm A)/M = (-i\hbar \bm \nabla - \bm A)/M$, with  $\bm j = \frac{1}{2}\left[ \Psi^\dagger \bm v\Psi +(\bm v\Psi)^\dagger\Psi\right]$.  In the present form, the particle current $\bm j$ in Eq.~(\ref{j})  appears to have a significant anisotropy owing to the contribution $ \hbar k_0\,\hat{\bm x}\,\cos \chi \,n_{TF}(r)$ from the vector potential, where the factor $\cos \chi$ arises from the expectation value of  $\sigma^z$ and $n_{TF}(r) =n_0(1-r^2/R^2)$ is the Thomas-Fermi particle density.

The evaluation of the first term in Eq.~(\ref{j}) yields the expected circulating flow around the vortex, but it also includes the gradient of the additional phase $\alpha x$ in Eq.~(\ref{phase2}), leading to the extra term $\alpha \,\hat{\bm x}\, \hbar \,n_{TF}(r)/M$.  Use of the optimal (minimal) value $\alpha =-k_0\cos \chi$ from Eq.~(\ref{alpha}) shows that this phase factor precisely cancels the explicit contribution from $\bm A$ in Eq.~(\ref{j}), leading to the  particle current expected for a  vortex in a two-component superfluid
\begin{equation}\label{jv}
\bm j(\bm r) = \frac{\hbar n_{TF}(r)}{2M}\,\frac{1}{2}\left[m_1+m_2 + (m_1-m_2)\cos\chi\right]\,\frac{\hat{\bm z}\times (\bm r - \bm r_0)}{|\bm r-\bm r_0|^2}.
\end{equation}

The present  uniform  Abelian  synthetic matrix gauge field $\bm A$ with vanishing synthetic magnetic field 
$\bm B = \bm \nabla\times \bm A = 0$ may represent a special case.  Here, the phase parameter $\alpha$ is effectively a gauge choice that can eliminate the uniform  $\bm A$.  It is not clear whether a similar cancelation can arise in more complicated geometries where $\bm \nabla\times \bm A$ is nonzero.

\section{Discussion and conclusions}

This study of vortex dynamics in a spin-orbit coupled Bose-Einstein condensate has relied on a variational time-dependent Lagrangian formalism.  In any such approach, the choice of various variational parameters determines the trial wave function,  and the resulting  solution represents the optimal description within the set of assumed parameters.

From this perspective, an improved set of parameters should yield different and more accurate dynamics.  
 Specifically, the spinor wave function in Eq.~(\ref{zeta})  assumes spatially constant parameters $\chi$ and $\eta$, which provides the variational solution for this particular trial function. Here, I choose them to minimize the energy in the absence of a vortex, but  they should, in principle, include the presence of the vortex and vary spatially. This approximation relies on the difference in energy scales between the external laser beam (the recoil energy $E_R$) and the much smaller vortex energy $E_v$. In addition, the vortex position $\bm r_0$ is assumed to be the same for both components, which simplifies the algebra but may not be the optimal situation~\cite{Radi11}.  Finally, a more general choice of interaction parameters $g_{jk}$ may lead to different TF radii for the two condensates.  Any or all of these additional effects are probably best included in a full numerical solution (I anticipate such studies in the foreseeable future). 

It is clear from Eqs.~(\ref{phidot}) and (\ref{udot}) that the Thomas-Fermi density profile leads to singular vortex dynamics near the TF boundary.  A more detailed numerical analysis of the two-component coupled GP equations should give an improved description for the behavior near the outer boundary.  In addition, I assume the $\hbar\Omega/E_R\lesssim 0.2$ to ensure miscibility of the two components~\cite{Lin11,Lu13}.  It would be interesting to have experiments for larger Rabi frequency~$\Omega$.

The single-component thermal-quench experiment~\cite{Frei10} uses a magnetic harmonic trap that allows the transition from the bound hyperfine state to one that is not bound.  Such a trap is probably infeasible for a spin-orbit coupled system that requires an external magnetic field to split the Zeeman hyperfine levels.  Reference~\cite{Rama12} studies the  corresponding situation in an optical trap and shows that it remains possible to out-couple a small fraction of the condensate to provide successive images, presumably even for a spin-orbit coupled BEC.
 
\section*{Acknowledgement} I am grateful to the Aspen Center for Physics and the NSF Grant \#1066293 for hospitality where part of this work was performed, including extended discussions with D.\ E.\ Sheehy.  In addition, I thank  S.~Reimann for hosting a visit to Lund University, where many of   these ideas were clarified.   J.\  Berlinsky, C.\ Kallin, G.~Kavoulakis,  B.\ L.\ Lev, P.\ Mason, and S.\  Reimann provided 
valuable comments that are much appreciated.  Finally, S.~Lederer, A.~Linde,  L.~Peeters, and Y.-F.\  Gu
 helped with formatting the figures, and I thank them for their valuable assistance.

\end{document}